\documentclass[preprint]{aastex}
\usepackage{graphicx}
\newcommand{\mjup}{M$_{\mathrm{Jup}}$}
\newcommand{\Teff}{$T_{\mathrm{eff}}$}
\slugcomment{To be submitted to \apj .}

\shorttitle{L' Exoplanet Survey of Sunlike Stars}
\shortauthors{Heinze et al.}

\begin{document}

\title{Constraints on Long-Period Planets from an $L'$ and
$M$ band Survey of Nearby Sun-Like Stars: Modeling Results\altaffilmark{1}}

\author{A. N. Heinze}
\affil{Steward Observatory, University of Arizona, 933 N Cherry Avenue, Tucson, AZ 85721}
\email{ariheinze@hotmail.com}
\author{Philip M. Hinz}
\affil{Steward Observatory, University of Arizona, 933 N Cherry Avenue,
  Tucson, AZ 85721}
\email{phinz@as.arizona.edu}
\author{Matthew Kenworthy}
\affil{Steward Observatory, University of Arizona, 933 N Cherry Avenue,
  Tucson, AZ 85721}
\email{mkenworthy@as.arizona.edu}
\author{Michael Meyer}
\affil{Department of Physics, Swiss
Federal Institute of Technology (ETH-Zurich), ETH H\'{o}nggerberg,
CH-8093 Zurich, Switzerland}
\email{mmeyer@phys.ethz.ch}
\author{Suresh Sivanandam}
\affil{Steward Observatory, University of Arizona, 933 N Cherry Avenue,
  Tucson, AZ 85721}
\email{suresh@as.arizona.edu}
\author{Douglas Miller}
\affil{Steward Observatory, University of Arizona, 933 N Cherry Avenue,
  Tucson, AZ 85721}
\email{dlmiller@as.arizona.edu}

\altaffiltext{1}{Observations reported here were obtained at the MMT
Observatory, a joint facility of the University of Arizona and the
Smithsonian Institution.}

\begin{abstract}
We have carried out an $L'$ and $M$ band Adaptive Optics (AO) 
extrasolar planet imaging survey of 54 nearby, sunlike stars 
using the Clio camera at the MMT.  
Our survey concentrates more strongly than
all others to date on very nearby F, G, and K stars, in
that we have prioritized proximity higher than youth. Our
survey is also the first to include extensive observations in the $M$
band, which supplemented the primary $L'$ observations.
These longer wavelength bands are most useful for very nearby systems in
which low temperature planets with red IR colors (i.e. $H - L'$,
$H - M$) could be detected.  
The survey detected no planets, but
set interesting limits on planets and brown dwarfs in the star
systems we investigated.  We have interpreted our null result
by means of extensive Monte Carlo simulations, and constrained
the distributions of extrasolar planets in mass $M$ and
semimajor axis $a$.  If planets are distributed according
to a power law with $dN \propto M^{\alpha} a^{\beta} dM da$, normalized to be
consistent with radial velocity statistics, we find that a distribution
with $\alpha = -1.1$ 
and $\beta = -0.46$, truncated
at 110 AU, is ruled out at the 90\% confidence level. 
These particular values of $\alpha$ and $\beta$ are significant because
they represent the most planet-rich case consistent with
current statistics from radial velocity observations.
With 90\% confidence no more than 8.1\% of stars
like those in our survey have systems with three widely
spaced, massive planets like the A-star HR 8799.  Our observations 
show that giant planets in long-period orbits around sun-like stars
are rare, confirming the results of shorter-wavelength 
surveys, and increasing the robustness of the conclusion.

\end{abstract}


\section{Introduction}

Nearly 400 extrasolar planets have now been discovered
using the radial velocity (RV) method.  RV surveys currently
have good statistical completeness only for planets with 
periods of less than ten years \citep{cumming,butlercat,carnp}, 
due to the limited temporal baseline of the observations, 
and the need to observe for a complete orbital period to 
confirm the properties of a planet with confidence. 
The masses of discovered planets range from
just a few Earth masses \citep{hotNep} up to around 20 Jupiter
masses (\mjup).  We note that a 20 \mjup~object would be considered
by many to be a brown dwarf rather than a planet, but that there is 
no broad consensus on how to define the upper mass limit for
planets.  For a good overview of RV planets to date, see
\citet{butlercat} or \url{http://exoplanet.eu/catalog-RV.php}.

The large number of RV planets makes it possible to examine
the statistics of extrasolar planet populations.  Several
groups have fit approximate power law distributions in
mass and semimajor axis to the set of known extrasolar
planets (see for example \citet{cumming}).  Necessarily, 
however, these power laws are
not subject to observational constraints at orbital
periods longer than 10 years -- and it is at these
orbital periods that we find giant planets in our own
solar system.  We cannot obtain a good
understanding of planets in general without information
on long period extrasolar planets.
Nor can we see how our own solar system fits into
the big picture of planet formation in the galaxy
without a good census of planets in Jupiter- and
Saturn-like orbits around other stars.

Repeatable detections of extrasolar planets (as opposed
to one-time microlensing detections)
have so far been made by transit detection (e.g. \citet{hd209458}), by RV
variations \citep{51peg}, by astrometric wobble \citep{benedict}, 
or by direct imaging \citep{hr8799}.
Of these methods, transits are efficient only for 
detecting close-in planets.  As noted above, precision
RV observations have not been going on long enough to 
detect more than a few planets with periods longer than ten years, 
but even as RV temporal baselines increase, long period
planets will remain harder to detect due to their slow orbital velocities.
The amplitude of a star's astrometric wobble increases 
with the radius of its planet's orbit, but decades-long
observing programs are still needed to find long-period planets.  
Direct imaging is the only method that allows us to characterize
long-period extrasolar planets on a timescale of months 
rather than years or decades.

Direct imaging of extrasolar planets is technologically
possible at present only in the infrared, based on
the planets' own thermal luminosity, not on reflected
starlight.  The enabling technology is adaptive optics (AO),
which allows 6-10m ground-based telescopes to obtain diffraction
limited IR images several times sharper than those from
HST, despite Earth's turbulent atmosphere.  Theoretical
models of giant planets indicate that
such telescopes should be capable of detecting self-luminous
giant planets in large orbits around young, nearby stars.
The stars should be young because the glow of
giant planets comes from gravitational potential
energy converted to heat in their formation and
subsequent contraction: lacking any internal fusion,
they cool and become fainter as they age.

Several groups have published the results of AO imaging
surveys for extrasolar planets around F, G, K, or M stars
in the last five years (see for 
example \citet{masciadri,kasper,biller1,GDPS}; and \citet{chauvin}).  
Of these, most have used wavelengths in the 1.5-2.2 $\mu$m
range, corresponding to the astronomical $H$ and $K_S$
filters \citep{masciadri,biller1,GDPS,chauvin}.  They have targeted
mainly very young stars.  Because young stars are rare, the median
distance to stars in each of these surveys has been
more than 20 pc.

In contrast to those above, our survey concentrates on 
very nearby F, G, and K stars, with proximity prioritized more than 
youth in the sample selection.  The median distance to our
survey targets is only 11.2 pc.  Ours is also the first survey 
to include extensive observations in the $M$
band, and only the second to search solar-type stars in the $L'$
band (the first was \citet{kasper}).  The distinctive focus on older, very
nearby stars for a survey using longer wavelengths 
is natural: longer wavelengths
are optimal for lower temperature planets which are most likely
to be found in older systems, but which would be undetectable
around all but the nearest stars.  More information on our
sample selection, observations, and data analysis can be found
in our Observations paper, \citet{obspaper}, which also details 
our careful evaluation of our survey's sensitivity, including extensive
tests in which fake planets were randomly placed in the raw
data and then recovered by an experimenter who knew neither
their positions nor their number.  Such tests are essential
for establishing the true relationship between source
significance (i.e. 5$\sigma$, 10$\sigma$, etc.) and survey
completeness.

Our survey places constraints on a more mature population of 
planets than those that have focused on very young stars, 
and confirms that a paucity of giant planets
at large separations from sun-like stars is robustly 
observed at a wide range of wavelengths.

In Section \ref{sec:rv}, we review power law fits to the distribution
of known RV planets, including the normalization of the power laws.
In Section \ref{sec:tmod}, we present the constraints our survey
places on the distribution of extrasolar giant planets, based on
extensive Monte Carlo simulations.  In Section \ref{sec:long} we
discuss the promising future of planet-search observations in the
$L'$ and especially the $M$ band, and in Section \ref{sec:concl}
we conclude.

\section{Statistical Distributions from RV Planets} \label{sec:rv}

Nearly 400 RV planets are known. See \citet{butlercat}
for a useful, conservative listing of confirmed extrasolar
planets as of 2006, or \url{http://exoplanet.eu/catalog-RV.php}
for a frequently-updated catalog of all confirmed and many
suspected extrasolar planet discoveries.  

The number of
RV planets is sufficient for meaningful statistical
analysis of how extrasolar planets are distributed in
terms of their masses and orbital semimajor axes.  The lowest mass 
planets and those with the longest orbital periods are
generally rejected from such analyses to reduce bias from
completeness effects, but there remains a considerable
range (2-2000 days in period, or roughly 0.03-3.1 AU
in semimajor axis for solar-type stars; and 0.3-20 \mjup~in mass)
where RV searches have good completeness \citep{cumming}.  There is evidence
that the shortest period planets, or `hot Jupiters,'
represent a separate population, a `pileup' of planets
in very close-in orbits that does not follow the same
statistical distribution as planets in more distant
orbits \citep{cumming}.  The hot Jupiters are therefore
often excluded from statistical fits to the overall
populations of extrasolar planets, or at least from the fits
to the semimajor axis distribution.

\citet{cumming} characterize the distribution of
RV planets detected in the Keck Planet Search
with an equation of the form 

\begin{equation}
 dN =C_0 M^{\alpha_L} P^{\beta_L} d\ln(M) d\ln(P).
\label{eq:cumming}
\end{equation}

where $M$ is the mass of the planet, $P$ is the orbital
period, and $C_0$ is a normalization constant.  They
state that 10.5\% of solar-type stars have a planet
with mass between 0.3 and 10 \mjup~and period between
2 and 2000 days, which information can be used to
derive a value for $C_0$ given values for the power
law exponents $\alpha_L$ and $\beta_L$.  They find
that the best-fit values for these are $\alpha_L = -0.31 \pm 0.2$
and $\beta_L = 0.26 \pm 0.1$, where the $_L$ subscript is our
notation to make clear that these are the exponents
for the form using logarithmic differentials.

In common with a number of other groups, we 
choose to represent the power law with ordinary differentials, 
and to give it in terms of orbital semimajor axis $a$ rather than
orbital period $P$:

\begin{equation}
 dN =C_0 M^{\alpha} a^{\beta} dM da.
\label{eq:uspower}
\end{equation}

Where $C_0$, of course, will not generally have
the same value for Equations \ref{eq:cumming} and
\ref{eq:uspower}.  Manipulating the two equations and using Kepler's
Third Law makes it clear that

\begin{equation}
\alpha = \alpha_L - 1.
\label{eq:setalpha}
\end{equation}

and

\begin{equation}
\beta = \frac{3}{2}\beta_L - 1.
\label{eq:setbeta}
\end{equation}

The \citet{cumming} exponents produce $\alpha=-1.31 \pm 0.2$
and $\beta = -0.61 \pm 0.15$ when translated into our form.
The mass power law is well behaved, but the integral of the
semimajor axis power law does not converge as $a \rightarrow \infty$,
so an outer truncation radius is an important parameter
of the semimajor axis distribution.

\citet{butlercat} presents the 2006 Catalog of Nearby 
Exoplanets, a carefully described heterogenous sample 
of planets detected by several different RV search programs.  
With appropriate caution, \citet{butlercat} refrain from quoting confident
power law slopes based on the combined discoveries
of many different surveys with different detection limits
and completeness biases (in contrast, the \citet{cumming}
analysis was restricted to stars in the Keck Planet
Search, which were uniformly observed up to a given minimum
baseline and velocity precision).  \citet{butlercat} do tentatively adopt
a power law with the form of Equation \ref{eq:uspower}
for mass only, and state that $\alpha$ appears to be
about -1.1 (or -1.16, to give the exact result of a formal
fit to their list of exoplanets).  However they caution
that due to their heterogeneous list of planets discovered
by different surveys, this power law should be taken more as a descriptor
of the known planets than of the underlying distribution.
They do not quote a value for the semimajor axis power
law slope $\beta$.

Based mostly on \citet{cumming}, but considering \citet{butlercat}
as helpful additional input, we conclude that the true
value of the mass power law slope $\alpha$ is probably between
-1.1 and -1.51, with -1.31 as a good working model.  The
value of the semimajor axis power law slope $\beta$ is probably
between -0.46 and -0.76, with -0.61 as a current best guess.  
The outer truncation radius of
the semimajor axis distribution cannot be constrained by
the RV results: surveys like ours exist, in part, to
constrain this interesting number.

The only other result we need from the RV searches
is a normalization that will allow us to find $C_0$.
We elect not to use the \citet{cumming} value
(10.5\% of stars having a planet
with mass between 0.3 and 10 \mjup~and period between
2 and 2000 days), because this range includes the
hot Jupiters, a separate population.  

We take our normalization instead from the Carnegie
Planet Sample, as described in \citet{carnp}.
Their Table 1 (online only) lists 850 
stars that have been thoroughly investigated with RV.
They state that all planets with mass at 
least 1 \mjup~and orbital period less
than 4 years have been detected around these stars.  
Forty-seven of these stars are marked in
Table 1 as having RV planets.  Table 2 from \citet{carnp}
gives the measured properties
of 124 RV planets, including those orbiting 45 of the
47 stars listed as planet-bearing in Table 1.  The stars 
left out are HD 18445 and and HD 225261.  We cannot
find any record of these stars having planets, and therefore
as far as we can tell they are typos in Table 1.  

Since all planets with masses above 1 \mjup~and periods less
than 4 years orbiting stars in the \citet{carnp} list
of 850 may be relied upon to have been discovered, we
may pick any sub-intervals in this range of mass and
period, and divide the number of planets falling into
these intervals by 850 to obtain our normalization.
We selected the range 1-13 \mjup~in mass, and 0.3-2.5 AU
in semimajor axis.  Twenty-eight stars, or 3.29\% of
the 850 in the \citet{carnp} list, have one or more planets in this
range.  Our inner limit of 0.3 AU excludes the hot Jupiters, and
thus the 3.29\% value provides our final normalization.
We note that if we adopt the \citet{cumming} best-fit power
laws, and use the 3.29\% normalization to predict the percentage
of stars having planets with masses between 0.3 and 10
\mjup~and orbital periods between 2 and 2000 days, we find a value
of 9.3\%, which is close to the \citet{cumming} value
of 10.5\%.  The slight difference is probably not significant,
but might be viewed as upward bias in the \citet{cumming} value
due to the inclusion of the hot Jupiters.  In any case we would
not have obtained very different constraints if we had
used the \citet{cumming} normalization in our
Monte Carlo simulations.

For comparison, among the other papers reporting
Monte Carlo simulations similar to ours,
\citet{kasper} used a normalization of 3\% for
planets with semimajor axes of 1-3 AU and masses greater
than 1 \mjup.  This is close to our value of 3.29\% for a
similar range.  \citet{GDPS} and \citet{nielsen} fixed
$\alpha$ and $\beta$ in their simulations, and let
the normalization be a free parameter. \citet{chauvin}
obtained their normalization from \citet{cumming},
and \citet{nielsenclose} obtained theirs from \citet{carnp}.

\citet{juric} provide a helpful
mathematical description of the eccentricity distribution
of known RV planets:

\begin{equation}
P(\epsilon) = \epsilon e^{-\epsilon^2/(2\sigma^2)}.
\label{eq:juric}
\end{equation}

where $P(\epsilon)$ is the probability of a given extrasolar
planet's having orbital eccentricity $\epsilon$, $e$ is the 
root of the natural logarithm, and $\sigma = 0.3$.
We find that this mathematical form provides an excellent fit to
the distribution of real exoplanet eccentricities 
from Table 2 of \citet{carnp}, so we have used it
as our probability distribution to generate
random eccentricities for the Monte Carlo simulations we describe
in Section \ref{sec:tmod} below.

\section{Constraints on the Distribution of Planets} \label{sec:tmod}

\subsection{Theoretical Spectra}
\citet{bur} present high resolution, flux-calibrated theoretical spectra
of giant planets or brown dwarfs for ages ranging from
0.1-5.0 Gyr and masses from 1 to 20 \mjup~(these are available
for download from \url{http://www.astro.princeton.edu/\textasciitilde burrows/}).  We
have integrated these spectra to give absolute
magnitudes in the $L'$ and $M$ filters used in Clio
(see Tables \ref{tab:burl} and \ref{tab:burm}), and have found that
the results can be reasonably interpolated to give the $L'$ or
$M$ band magnitudes for all planets of interest for
our survey.  \citet{bar} also present models of giant
planets and brown dwarfs, pre-integrated into magnitudes
in the popular infrared bands.  These models predict
slightly better sensitivity to low mass planets in the $L'$ band
and slightly poorer sensitivity in the $M$ band, relative
to the \citet{bur} models.  We cannot say if the difference
is due to the slightly different filter sets used (MKO
for Clio vs. Johnson-Glass and Johnson for \citet{bar}),
or if it is intrinsic to the different model spectra
used in \citet{bur} and \citet{bar}.  We have chosen
to use the \citet{bur} models exclusively herein, to
avoid any errors due to the slight filter differences.  
Since the \citet{bur} models predict poorer
sensitivity in the $L'$ band, in which the majority
of our survey was conducted, our decision to use
them is conservative.

\begin{deluxetable}{cccccc}
\tablewidth{0pc}
\tablecolumns{6}
\tablecaption{$L'$ Band Absolute Mags from \citet{bur} \label{tab:burl}}
\tablehead{\colhead{Planet Mass} & \colhead{Mag at} & \colhead{Mag at} & \colhead{Mag at} & \colhead{Mag at} & \colhead{Mag at}\\
\colhead{in \mjup} & \colhead{0.10 Gyr} & \colhead{0.32 Gyr} & \colhead{1.0 Gyr} & \colhead{3.2 Gyr} & \colhead{5.0 Gyr}} 
\startdata
\phantom{0o0} 1.0 \phantom{0o0} & \phantom{0o0} 19.074 \phantom{0o0} & \phantom{0o0} 23.010 \phantom{0o0} & \phantom{0o0} 27.870 \phantom{0o0} & \phantom{0o0} 33.50\tablenotemark{a} \phantom{0o0} & \phantom{0o0} 35.50\tablenotemark{a} \phantom{0o0} \\
2.0 & 16.793 & 19.351 & 23.737 & 28.398 & 29.479 \\
5.0 & 14.500 & 16.397 & 18.588 & 22.437 & 24.407 \\
7.0 & 13.727 & 15.390 & 17.336 & 20.131 & 21.574 \\
10.0 & 12.888 & 14.437 & 16.246 & 18.480 & 19.466 \\
15.0 & 12.00\tablenotemark{b} & 13.61\tablenotemark{b} & 14.773 & 16.816 & 17.691 \\
20.0 & 11.30\tablenotemark{b} & 12.98\tablenotemark{b} & 14.190 & 15.967 & 16.766 \\
\enddata
\tablenotetext{a}{\footnotesize{No models for these
very faint planets appear in \citet{bur}. We have inserted ad hoc 
values to smooth the interpolations.  Any effect of
the interpolated magnitudes for planets we could actually detect
is negligible.}}
\tablenotetext{b}{\footnotesize{No models for these
bright, hot planets appear in \citet{bur}, which focuses
on cooler objects.  We have added values from \citet{bar} and then 
adjusted them to slightly fainter values to ensure smooth interpolations.}}
\end{deluxetable}

\begin{deluxetable}{cccccc}
\tablewidth{0pc}
\tablecolumns{6}
\tablecaption{$M$ Band Absolute Mags from \citet{bur} \label{tab:burm}}
\tablehead{\colhead{Planet Mass} & \colhead{Mag at} & \colhead{Mag at} & \colhead{Mag at} & \colhead{Mag at} & \colhead{Mag at}\\
\colhead{in \mjup} & \colhead{0.10 Gyr} & \colhead{0.32 Gyr} & \colhead{1.0 Gyr} & \colhead{3.2 Gyr} & \colhead{5.0 Gyr}} 
\startdata
\phantom{0o0} 1.0 \phantom{0o0} & \phantom{0o0} 14.974 \phantom{0o0} & \phantom{0o0} 16.995 \phantom{0o0} & \phantom{0o0} 19.987 \phantom{0o0} & \phantom{0o0} 25.0\tablenotemark{a} \phantom{0o0} & \phantom{0o0} 26.0\tablenotemark{a} \phantom{0o0} \\
2.0 & 14.023 & 15.313 & 17.807 & 21.295 & 22.163 \\
5.0 & 13.014 & 14.017 & 15.153 & 17.167 & 18.537 \\
7.0 & 12.618 & 13.561 & 14.558 & 16.126 & 16.909 \\
10.0 & 12.189 & 13.096 & 14.093 & 15.315 & 15.951 \\
15.0 & 11.55\tablenotemark{b} & 12.60\tablenotemark{b} & 13.370 & 14.512 & 14.990 \\
20.0 & 11.29\tablenotemark{b} & 12.21\tablenotemark{b} & 13.069 & 14.122 & 14.580 \\
\enddata
\tablenotetext{a}{\footnotesize{No models for these
very faint planets appear in \citet{bur}. We have inserted ad hoc 
values to smooth the interpolations.  Any effect of
the interpolated magnitudes for planets we could actually detect
is negligible.}}
\tablenotetext{b}{\footnotesize{No models for these
bright, hot planets appear in \citet{bur}, which focuses
on cooler objects.  We have added values from \citet{bar} and then 
adjusted them to slightly fainter values to ensure smooth interpolations.}}
\end{deluxetable}

\subsection{Introducing the Monte Carlo Simulations} \label{sec:mcint}
In common with several other surveys \citep{kasper,biller1,GDPS,chauvin}
we have used our survey null result to set upper limits on planet
populations via Monte Carlo simulations.  In these simulations,
we input our sensitivity data in the form of tabular files giving
the sensitivity in apparent magnitudes as a function of separation
in arcseconds for each star.  Various features of our images
could cause the sensitivity at a given separation to vary somewhat
with position angle: to quantify this, our tabular files give
ten different values at each separation, corresponding
to ten different percentiles ranging from the worst to the best
sensitivity attained at that separation.  These files are described in detail
in \citet{obspaper}, and are available for download
from \url{http://www.hopewriter.com/Astronomyfiles/Data/SurveyPaper/}

The Monte Carlo simulations described below allow us to use the
observed sensitivity to planets in our survey to calculate directly
the probability of a given parameter or set of parameters describing
the exoplanet population.  This, in turn, allows us to constrain these
 parameters at a given confidence level.  This is a maximum likelihood
technique that allows us to incorporate all the individual probability
functions of the data, as well as parameterized models of the
exoplanet population.  The approach is similar to a Bayesian approach.
 However, we also use the results of the simulations to set confidence
limits to the parameters, a more classical approach.

Each Monte Carlo simulation runs with given planet distribution
power law slopes $\alpha$ and $\beta$, and a given outer 
truncation value $R_{trunc}$
for the semimajor axis distribution.  Using the normalization 
described in Section \ref{sec:rv}, the probability $P_{plan}$ 
of any given star having a planet
between 1 and 20 \mjup~is then calculated from the 
input $\alpha$, $\beta$, and $R_{trunc}$.  In each
realization of our survey, each star is randomly assigned 
a number of planets, based on Poisson statistics 
with mean $P_{plan}$.  In most cases $P_{plan} << 1$,
so the most likely number of planets is zero.  If the 
star turns out to have one or more planets, the mass 
and semimajor axis of each are randomly selected
from the input power law distributions. The eccentricity 
is randomly selected from the
\citet{juric} distribution, and an inclination is 
randomly selected from the distribution $P(i) \propto \sin(i)$. 
If the star is a binary, the planet may be dropped from 
the simulation at this point if the orbit seems likely 
to be unstable.  In general, we consider circumstellar
planets to be stable as long as their apastron distance is less than
$1/3$ the projected distance to the companion star, and circumbinary
planets to be stable as long as their periastron distance
is at least three times greater than the projected separation 
of the binary.  For planets orbiting low-mass secondaries, 
a smaller limit on the apastron distance is sometimes
imposed, while often circumbinary planets required such distant
orbits that they were simply not considered; the details are given
in Table \ref{tab:binaries}.  For each planet passing the orbital
stability checkpoint, a full orbit is calculated
using a binary star code written by one of us (M. K.).  
The projected separation
in arcseconds is found, and the magnitude of the planet is calculated from its
mass, distance, and age using the \citet{bur} models.

Two further random choices complete the determination of whether
the simulated planet is detected.  First, one of the ten percentiles
given in the sensitivity files is randomly selected.  Combined
with the separation in arcseconds, this selection specifies
the sensitivity of our observation at the location of the simulated
planet.  The second random choice is needed because planets
appearing at low significance in our images would have
a less than 100\% chance of being confidently detected. 
Our blind sensitivity tests using fake planets placed in our 
raw data showed that we could confirm 97\% of 10$\sigma$ sources,
46\% of 7$\sigma$ sources, and 16\% of 5$\sigma$ sources, 
where $\sigma$ is a measure of the PSF-scale noise in a 
given region of the image (see \citet{obspaper} for details). 
This second and final random choice in our Monte Carlo simulations
is therefore arranged to ensure that
a randomly selected 16\% of planets with 5-7$\sigma$ significance,
and 46\% of planets with 7-10$\sigma$ significance,
are recorded in the simulation as detected objects.  
Although we have 97\% completeness at 10$\sigma$, we choose
to consider 100\% of simulated planets with 10$\sigma$ or 
greater significance to be detected, because at only slightly
above 10$\sigma$ the true completeness certainly becomes 100\% for
all practical purposes.  Note that we have conservatively
allowed the detection probabilities to increase stepwise,
rather than in a continuous curve, from 5 to 10$\sigma$: that is,
in our Monte Carlo simulations, planets with 5-7$\sigma$ significance
are detected at the 5$\sigma$ rate from our blind sensitivity
tests, while those with 7-10$\sigma$ significance are detected
at the 7$\sigma$ rate.

The low completeness (16\%) at 5$\sigma$, as determined from 
our blind sensitivity tests using fake planets, may seem surprising.
In these tests we distinguished between planets that were suggested by
a concentration of unusually bright pixels (`Noticed'), or else
confidently identified as real sources (`Confirmed').
Many more planets were noticed
than were confirmed: for noticed planets, the rates
are 100\% at 10$\sigma$, 86\% at 7$\sigma$, and 56\% at 5$\sigma$.
However, very many false positives were also noticed, so
sources that are merely noticed but not confirmed do not
represent usable detections.  The completeness levels
we used in our Monte Carlo simulations (16\% at 5$\sigma$ and
46\% at 7$\sigma$) refer to confirmed
sources.  No false positives were confirmed in any of our 
blind tests.  Followup observations of suspected sources are costly in terms of
telescope time, so a detection strategy with a low false-positive
rate is important.  

Though sensitivity estimators (and therefore
the exact meaning of 5$\sigma$) differ among
planet imaging surveys, ours was quite
conservative, as is explained in \citet{obspaper}.  
The low completeness we find at 5$\sigma$, which
has often been taken as a high-completeness sensitivity
limit, should serve as a warning to future workers in this field, and an
encouragement to establish a definitive significance-completeness
relation through blind sensitivity tests as 
we have done.

Note that our blind sensitivity tests,
covered in \citet{obspaper}, are completely distinct
from the Monte Carlo simulations covered herein.  The blind
tests involved inserting a little over a hundred fake planets
into our raw image data to establish our point-source sensitivity.  
In our Monte Carlo work we simulated the orbits, masses,
and brightnesses of millions of planets, and compared
them to our previously-established sensitivity limits
to see which planets our survey could have detected.

\subsection{A Detailed Look at a Monte Carlo Simulation} \label{sec:mcdet}

To evaluate the significance of our survey and provide some
guidance for future work, we have analyzed in detail a single Monte
Carlo simulation.  We chose the \citet{cumming} best fit
values of $\alpha = -1.31$ and $\beta = -0.61$, with
the semimajor axis truncation radius set to 100 AU.  
Planets could range in mass from 1 to 20 \mjup.  
As described in Section \ref{sec:rv} above, we normalized
the planet distributions so that each star had a 3.29\% probability
of having a planet with semimajor axis between 0.3 and 2.5
AU and mass between 1 and 13 \mjup.  
The simulation consisted of 50,000 realizations of our survey
with these parameters.  In all, 505,884 planets were
simulated, of which 51,879 were detected.  

In 38\% of the 50,000 realizations, our survey found zero planets,
while 37\% of the time it found one, and 25\% of the time it found
two or more.  The planet 
distribution we considered in this simulation
cannot be ruled out by our survey, since a null result
such as we actually obtained turns out not to be very improbable.

\begin{figure}
\plotone{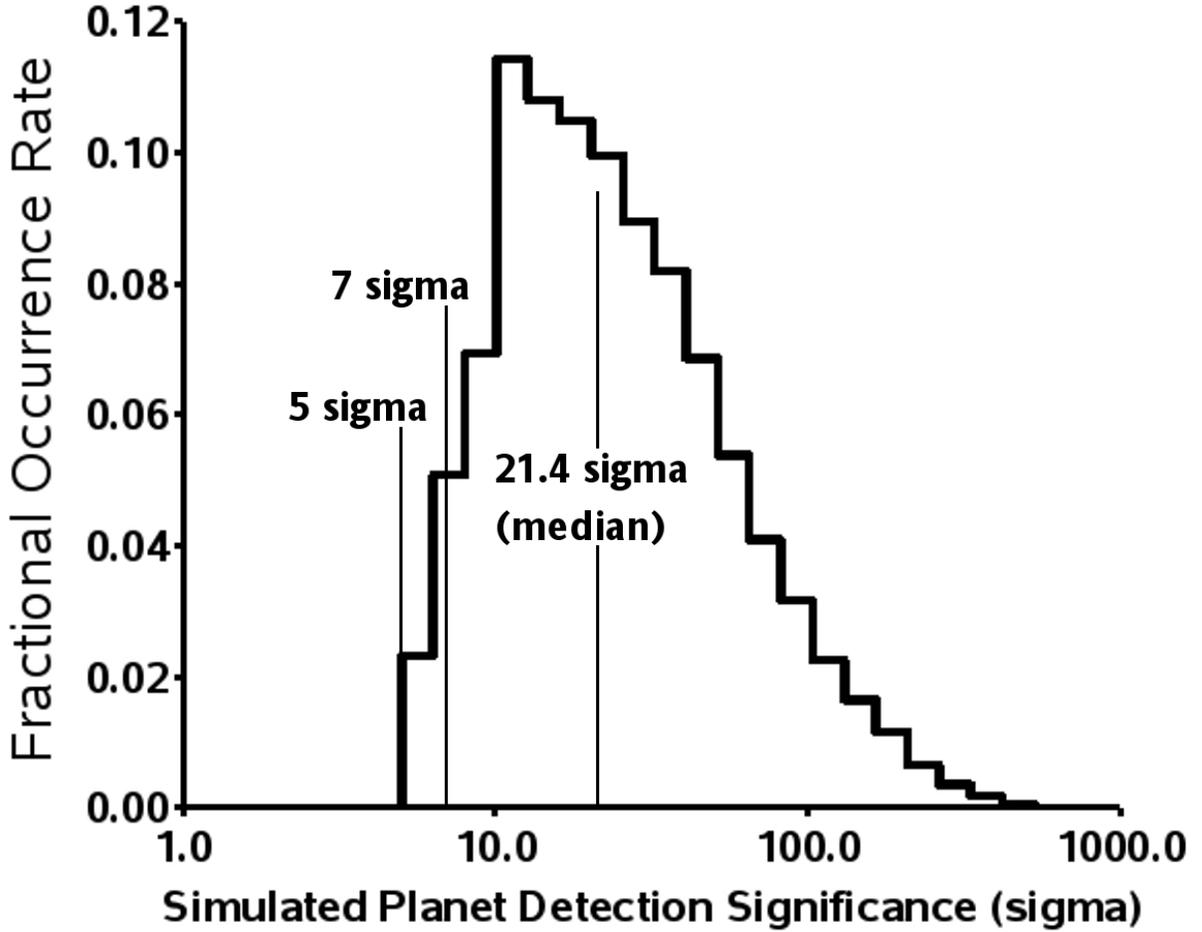}
\caption{Histogram of detection significance for the
51,879 simulated planets detected in 50,000 realizations of our
survey with the \citet{cumming} distribution ($\alpha = -1.31$,
$\beta = -0.61$) truncated at 100 AU.  Our detection
rates went down for significance less than 10 $\sigma$,
but some 5-7 $\sigma$ planets are still detected.  The
relatively high median significance of 21.4 $\sigma$ suggests
any detected planet would most likely be quite obvious --
a good argument for doing `quick-look' data reductions
as soon as possible at the telescope.
\label{fig:sighist}}
\end{figure}

The large number of survey realizations in our simulation allows the
calculation of precise statistics for potentially detectable planets.
The median mass of detected planets in our simulation was 11.36 \mjup, 
the median semimajor axis was 43.5 AU, the median angular 
separation was 2.86 arcsec,
and the median significance was 21.4$\sigma$.  This last number
is interesting because it suggests that, for our
survey, any real planet detected was likely to appear at high
significance, obvious even on a preliminary, `quick-look' 
reduction of the data.  This suggests that performing such 
reductions at the telescope should be a high priority, to 
allow immediate confirmation and followup if a candidate
is seen.  Figure \ref{fig:sighist} presents as a histogram
the significance of all planets detected in this Monte Carlo simulation.

We suspected that there would be a detection bias
toward very eccentric planets, because these would spend
most of their orbits near apastron, where they would be
easier to detect.  This bias did not appear at any
measurable level in our simulation.  However, there was
a weak but clear bias toward planets in low-inclination
orbits, which, of course, spend more of their time
at large separations from their stars than do planets
with nearly edge-on orbits.

A concern with any planet imaging survey is how strongly the results
hinge on the best (i.e. nearest and youngest) few stars.  
A survey of 54 stars may have far less
statistical power than the number would imply if the best two or three
stars had most of the probabilty of hosting detectable planets.
Table \ref{tab:sdp} gives the percentage of planets detected
around each star in our sample based on our detailed Monte Carlo
simulation.  Due to poor data quality, binary orbit constraints,
or other issues, a few stars had zero probability of detected planets
given the distribution used here.  In general, however, the likelihood of 
hosting detectable planets is fairly well distributed.  

In Table \ref{tab:binaries}, we give the details of planetary
orbital constraints used in our Monte Carlo simulations for
each binary star we observed, complete with the separations
we measured for the binaries.  Note that HD 96064 B
is a close binary star in its own right, so planets orbiting
it were limited in two ways: the apastron could not be too
far out, or the orbit would be rendered unstable by proximity
to HD 96064 A -- but the periastron also could not be too far in,
or the binary orbit of HD 96064 Ba and HD 96064 Bb would render
it unstable.  Planets individually orbiting 
HD 96064 Ba or HD 96064 Bb were not considered in our survey,
since to be stable the planets would have to be far too close-in 
for us to detect them.  The constraints described
in Table \ref{tab:binaries} account for most of the stars in
Table \ref{tab:sdp} with few or no detections reported.

A final question our detailed simulation can address
is how important the $M$ band observations
were to the survey results.  In Table \ref{tab:mband}, we show
that when $M$ band observations were made, they 
did substantially increase
the number of simulated planets detected.

\begin{center}
\begin{deluxetable}{lcccc}
\tablewidth{0pc}
\tablecolumns{5}
\tablecaption{Percentage of Detected Planets Found Around Each Star \label{tab:sdp}}
\tablehead{ & \colhead{\% of Total} & \colhead{Median} & \colhead{Median} &
  \colhead{Median} \\
\colhead{Star Name} & \colhead{Detected Planets} & \colhead{Mass} &
\colhead{Semimajor Axis} & \colhead{Separation}}
\startdata
GJ 117 & 6.07 & 7.66 \mjup~& 39.36 AU & 3.64 arcsec \\
$\epsilon$ Eri & 5.83 & 6.98 \mjup~& 18.26 AU & 4.35 arcsec \\
HD 29391 & 5.80 & 8.14 \mjup~& 49.13 AU & 2.71 arcsec \\
GJ 519 & 4.74 & 10.44 \mjup~& 40.51 AU & 3.28 arcsec \\
GJ 625 & 4.67 & 9.72 \mjup~& 29.18 AU & 3.48 arcsec \\
GJ 5 & 4.45 & 9.60 \mjup~& 53.42 AU & 3.08 arcsec \\
BD+60 1417 & 3.95 & 11.58 \mjup~& 44.48 AU  & 2.05 arcsec \\
GJ 355 & 3.81 & 9.71 \mjup~& 53.91 AU & 2.34 arcsec \\
GJ 354.1 A & 3.67 & 9.58 \mjup~& 60.12 AU & 2.64 arcsec \\
GJ 159 & 3.57 & 9.73 \mjup~& 57.95 AU & 2.71 arcsec \\
GJ 349 & 3.35 & 11.38 \mjup~& 44.40 AU & 3.17 arcsec \\
61 Cyg B & 3.29 & 11.32 \mjup~& 19.53 AU & 4.08 arcsec \\
GJ 879 & 3.03 & 11.18 \mjup~& 36.84 AU & 3.69 arcsec \\
GJ 564 & 2.94 & 10.67 \mjup~& 56.80 AU & 2.70 arcsec \\
GJ 410 & 2.93 & 12.78 \mjup~& 41.83 AU & 3.03 arcsec \\
GJ 450 & 2.89 & 12.90 \mjup~& 38.72 AU & 3.66 arcsec \\
GJ 3860 & 2.68 & 12.70 \mjup~& 49.72 AU & 2.69 arcsec \\
HD 78141 & 2.58 & 12.47 \mjup~& 57.00 AU & 2.24 arcsec \\
BD+20 1790 & 2.51 & 12.14 \mjup~& 58.33 AU & 2.02 arcsec \\
GJ 278 C & 2.20 & 12.68 \mjup~& 54.56 AU & 3.04 arcsec \\
GJ 311 & 2.19 & 12.55 \mjup~& 52.07 AU & 3.20 arcsec \\
HD 113449 & 2.17 & 12.52 \mjup~& 59.31 AU & 2.29 arcsec \\
GJ 211 & 2.10 & 13.59 \mjup~& 50.51 AU & 3.30 arcsec \\
BD+48 3686 & 2.08 & 12.56 \mjup~& 55.05 AU & 2.01 arcsec \\
GJ 282 A & 2.05 & 13.39 \mjup~& 49.85 AU & 2.99 arcsec \\
GJ 216 A & 2.03 & 12.71 \mjup~& 42.98 AU & 4.21 arcsec \\
61 Cyg A & 1.97 & 13.70 \mjup~& 20.94 AU & 4.54 arcsec \\
HD 1405 & 1.54 & 13.13 \mjup~& 66.34 AU & 2.04 arcsec \\
HD 220140 A & 1.54 & 11.73 \mjup~& 36.85 AU & 1.73 arcsec \\
HD 96064 A & 1.49 & 12.63 \mjup~& 46.64 AU & 1.75 arcsec \\
HD 139813 & 1.43 & 14.33 \mjup~& 59.71 AU & 2.37 arcsec \\
GJ 380 & 0.92 & 15.76 \mjup~& 25.31 AU & 4.21 arcsec \\
GJ 896 A & 0.61 & 12.43 \mjup~& 6.47 AU & 0.98 arcsec \\
GJ 860 A & 0.38 & 11.58 \mjup~& 53.26 AU & 6.62 arcsec \\
$\tau$ Ceti & 0.38 & 17.19 \mjup~& 25.49 AU & 5.52 arcsec \\
GJ 896 B & 0.34 & 11.40 \mjup~& 6.78 AU & 1.14 arcsec \\
$\xi$ Boo B & 0.32 & 12.07 \mjup~& 8.25 AU & 1.36 arcsec \\
HD 220140 B & 0.28 & 12.04 \mjup~& 25.92 AU & 1.37 arcsec \\
$\xi$ Boo A & 0.24 & 12.89 \mjup~& 8.72 AU & 1.50 arcsec \\
GJ 659 B & 0.21 & 17.71 \mjup~& 62.54 AU & 2.81 arcsec \\
GJ 166 B & 0.17 & 16.12 \mjup~& 6.19 AU & 1.34 arcsec \\
GJ 684 A & 0.17 & 14.93 \mjup~& 85.98 AU & 4.87 arcsec \\
HD 96064 B & 0.13 & 14.43 \mjup~& 38.55 AU & 1.60 arcsec \\
GJ 505 B & 0.12 & 15.94 \mjup~& 17.11 AU & 1.61 arcsec \\
GJ 166 C & 0.10 & 15.56 \mjup~& 6.43 AU & 1.52 arcsec \\
GJ 505 A & 0.07 & 16.32 \mjup~& 18.08 AU & 1.75 arcsec \\
GJ 702 A & 0.02 & 15.90 \mjup~& 6.21 AU & 1.50 arcsec \\
GJ 684 B & None & NA & NA & NA \\
GJ 860 B & None & NA & NA & NA \\
GJ 702 B & None & NA & NA & NA \\
HD 77407 A & None & NA & NA & NA \\
GJ 659 A & None & NA & NA & NA \\
GJ 3876 & None & NA & NA & NA \\
HD 77407 B & None & NA & NA & NA \\
\enddata
\tablecomments{This table applies
to our detailed Monte Carlo simulation
with 50,000 survey realizations run using
$\alpha=-1.31$, $\beta=-0.61$, and semimajor
axis truncation radius 100 AU.  Of all the simulated
planets that were detected, we present here
the percentage that were found around each given star,
and the median mass, semimajor axis, and projected
separation for simulated planets found around each star.
The table thus indicates around which stars our
survey had the highest likelihood of detecting
a planet.  Many stars with poor likelihood are
binaries, with few stable planetary orbits possible.}
\end{deluxetable}
\end{center}

\begin{center}
\begin{deluxetable}{lccccc}
\tablewidth{0pc}
\tablecolumns{6}
\tablecaption{Constraints on Simulated Planet Orbits Around Binary Stars\label{tab:binaries}}
\tablehead{ &  & \colhead{constraints on} &
  \colhead{constraints on} & \colhead{constraints on} \\ 
 & \colhead{separation} & \colhead{circumprimary} &
  \colhead{circumsecondary} & \colhead{circumbinary} \\ 
\colhead{Star Name} & \colhead{(arcsec)} & \colhead{apastron} & 
\colhead{apastron} & \colhead{periastron}}
\startdata
HD 220140 AB & 10.828 & $<$3.61 asec (71.3 AU) & $<$2.17 asec (42.8 AU) & No Stable Orbits \\
HD 96064 AB &  11.628 & $<$3.88 asec (95.6 AU) & $<$2.33 asec (57.3 AU) & No Stable Orbits\\
HD 96064 Bab &  0.217 & No Stable Orbits & No Stable Orbits & $>$0.65 asec
(16.1 AU) \\
GJ 896 AB &     5.366 & $<$1.79 asec (11.8 AU) & $<$1.79 asec (11.8 AU) & No Stable Orbits \\
GJ 860 AB &     2.386 & $<$0.79 asec (3.17 AU)  & $<$0.60 asec (2.41 AU) &
$>$7.15 asec (28.7 AU) \\
$\xi$ Boo AB &  6.345 & $<$2.12 asec (14.2 AU) & $<$2.12 asec (14.2 AU) & No Stable Orbits \\
GJ 166 BC &     8.781 & $<$2.20 asec (10.6 AU) & $<$2.20 asec (10.6 AU) & No Stable Orbits \\
GJ 684 AB &     1.344 & $<$0.45 asec (6.34 AU) & $<$0.27 asec (3.80 AU) &
$>$4.03 asec (56.8 AU) \\
GJ 505 AB &     7.512 & $<$2.50 asec (29.8 AU) & $<$2.50 asec (29.8 AU) & No Stable Orbits \\
GJ 702 A &      5.160 & $<$1.76 asec (8.85 AU) & $<$1.32 asec (6.64 AU) &
$>$15.9 asec (79.7 AU) \\
HD 77407 AB &   1.698 & $<$0.57 asec (17.2 AU) & $<$0.34 asec (10.2 AU) &
$>$5.11 asec (153.7 AU) \\
\enddata
\tablecomments{Planets orbiting the primary in a binary star
were considered to be de-stabilized by the gravity of the
secondary if their apastron distance from the primary was
too large.  Similarly, planets orbiting the secondary
had to have small enough apastron distances to avoid
being de-stabilized by the primary.  Circumbinary planets
had to have a large enough periastron distance to avoid
being de-stabilized by the differing gravitation of the
two components of the binary.  Note that HD 96064B is
itself a tight binary star, so planets orbiting it
had both a minimum periastron and a maximum apastron.
Constraints are given in AU as well as arcseconds so constraints
can easily be compared with actual or hypothetical planetary systems.}
\end{deluxetable}
\end{center}

\begin{center}
\begin{deluxetable}{lcccc}
\tablewidth{0pc}
\tablecolumns{5}
\tablecaption{Importance of the $M$ Band Data \label{tab:mband}}
\tablehead{ & \colhead{Total simulated} &\colhead{2-band} & \colhead{$L'$-only} & \colhead{$M$-only}\\
\colhead{Star Name} & \colhead{detections} & \colhead{detections} & \colhead{detections} & \colhead{detections}}
\startdata
$\epsilon$ Eri & 2850 & 46.98\% & 8.28\% & 44.74\% \\
61 Cyg B & 1610 & 52.73\% & 1.55\% & 45.71\% \\
61 Cyg A & 965 & 63.01\% & 22.80\% & 14.20\% \\
$\xi$ Boo B & 157 & 61.15\% & 18.47\% & 20.38\% \\
$\xi$ Boo A & 115 & 60.00\% & 18.26\% & 21.74\% \\
GJ 702 A & 9 & 22.22\% & 0.00\% & 77.78\% \\
\enddata
\tablecomments{The usefulness of $M$ band
observations, based on our detailed Monte
Carlo simulation.  When $M$ band observations 
were made of a given star, they did substantially increase
the number of simulated planets detected around that star.  }
\end{deluxetable}
\end{center}

\subsection{Monte Carlo Simulations: Constraining the Power Laws} \label{sec:mcbig}

The planet distribution we used in the single Monte Carlo
simulation described above could not be ruled out by our survey.
To find out what distributions could be ruled out, we performed
Monte Carlo simulations assuming a large number of different
possible distributions, parametrized by the two power
law slopes $\alpha$ and $\beta$, and by the outer semimajor
axis truncation radius $R_{trunc}$.  
Regardless of the values of $\alpha$ and $\beta$, each simulation 
was normalized to match the RV statistics
of \citet{carnp}: any given star had 3.29\% probability of
hosting a planet with mass between 1 and 13 \mjup~and 
semimajor axis between 0.3 and 2.5 AU.  
The mass range for simulated planets was 1-20 \mjup.

We tested three
different values of $\alpha$: -1.1, -1.31, and -1.51, roughly
corresponding to the most optimistic permitted, the best
fit, and the most pessimistic permitted values from \citet{cumming}.
For each value of $\alpha$, we ran simulations spanning
a wide grid in terms of $\beta$ and $R_{trunc}$.  In constrast
to the extensive results described in Section \ref{sec:mcdet},
the only data saved for these simulations was the probability
of finding zero planets.  Since we did in fact obtain a null
result, distributions for which the probability of this was
sufficiently low can be ruled out.  

\begin{figure}
\includegraphics[scale=6.0]{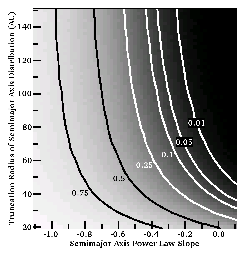}
\caption{Probability of our survey detecting zero planets,
as a function of the power law slope of the semimajor
axis distribution $\beta$, where $\frac{dn}{da} \propto a^{\beta}$,
and the outer truncation radius of the semimajor axis distribution.
Here, the slope of the mass distribution $\alpha$ has been taken as -1.31,
where $\frac{dn}{dM} \propto M^{\alpha}$.  Since we found
no planets, distributions that lead to a probability $P$ of
finding no planets are ruled out at the $1-P$ confidence
level: for example, the region above and to the right of the
0.1 contour is ruled out at the 90\% confidence level
\label{fig:bmc1.31}}
\end{figure}

\begin{figure}
\plottwo{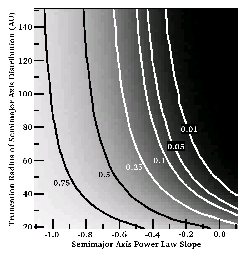}{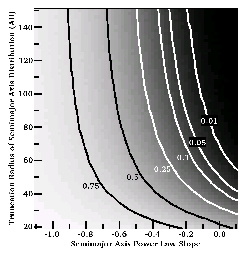}
\caption{Probability of our survey detecting zero planets,
as a function of the power law slope of the semimajor
axis distribution $\beta$, where $\frac{dn}{da} \propto a^{\beta}$,
and the outer truncation radius of the semimajor axis distribution.
Here, the slope of the mass distribution $\alpha$ has been taken as -1.1
(left) and -1.51 (right),
where $\frac{dn}{dM} \propto M^{\alpha}$.  Since we found
no planets, distributions that lead to a probability $P$ of
finding no planets are ruled out at the $1-P$ confidence
level: for example, the regions above and to the right of the
0.1 contours are ruled out at the 90\% confidence level
\label{fig:bmc_other}}
\end{figure}

Figures \ref{fig:bmc1.31} and \ref{fig:bmc_other} show the probability of a null
result as a function of $\beta$ and $R_{trunc}$ for our 
three different values of $\alpha$.  Figure \ref{fig:bmc1.31}
presents constraints based on $\alpha = -1.31$, the best-fit
value from RV statistics, while Figure \ref{fig:bmc_other}
compares the optimistic case $\alpha = -1.1$ and the
pessimistic case $\alpha = -1.51$.  Each pixel in these
figures represents a Monte Carlo simulation involving
15,000 realizations of our survey; generating the
figures took several tens of hours on a fast PC.
Contours are overlaid at selected probability
levels.  Regions within the 1\%, 5\%, and 10\% contours can, of course,
be ruled out at the 99\%, 95\%,
and 90\% confidence levels respectively.  For example,
we find that the most optimistic power laws allowed
by the \citet{cumming} RV statistics, $\alpha = -1.1$ and $\beta = -0.46$, 
are ruled out with 90\% confidence if $R_{trunc}$ is 110 AU
or greater.  Similarly, $\alpha = -1.51$ and $\beta = -0.3$, 
truncated at 100 AU, is ruled out.  Though
$\beta=0.0$ is not physically plausible, previous work
has sometimes used it as an example: for $\alpha=-1.31$,
we rule out $\beta=0.0$ unless $R_{trunc}$ is less than
38 AU.

\subsection{Model-independent Constraints} \label{sec:indep}

It is also possible to place constraints on the distribution
of planets without assuming a power law or any other particular
model for the statistics of planetary masses and orbits. Note
well that by ``model-independent'' in this context, we mean
independent only of models for the statistical distributions of
planets in terms of $M$ and $a$ -- not 
independent of models of planetary \textit{spectra} such 
those we obtain from \citet{bur}.  The latter are our only
means of converting from planetary mass and age to detectable flux,
and as such they remain indispensable.
 
To place our model-independent constraints, we performed an additional series
of Monte Carlo simulations on a grid of planet mass and
orbital semimajor axis.  For each grid point we seek to
determine a number $P(M,a)$ such that, with some specified
level of confidence (e.g., 90\%), the probability of a star like those
in our sample having a planet with the specified mass $M$
and semimajor axis $a$ is no more than $P(M,a)$.  We determine
$P(M,a)$ by a search: first a guess is made, and a Monte Carlo
simulation assuming this probability is performed.  If more
than 10\% of the realizations of our survey turn up a null
result, the guessed probability is too low; if less than 10\%
turn up a null result, the probability is too high.  It is
adjusted in steps of ever-decreasing size until the correct value
is reached.

\begin{figure}
\includegraphics[scale=6.0]{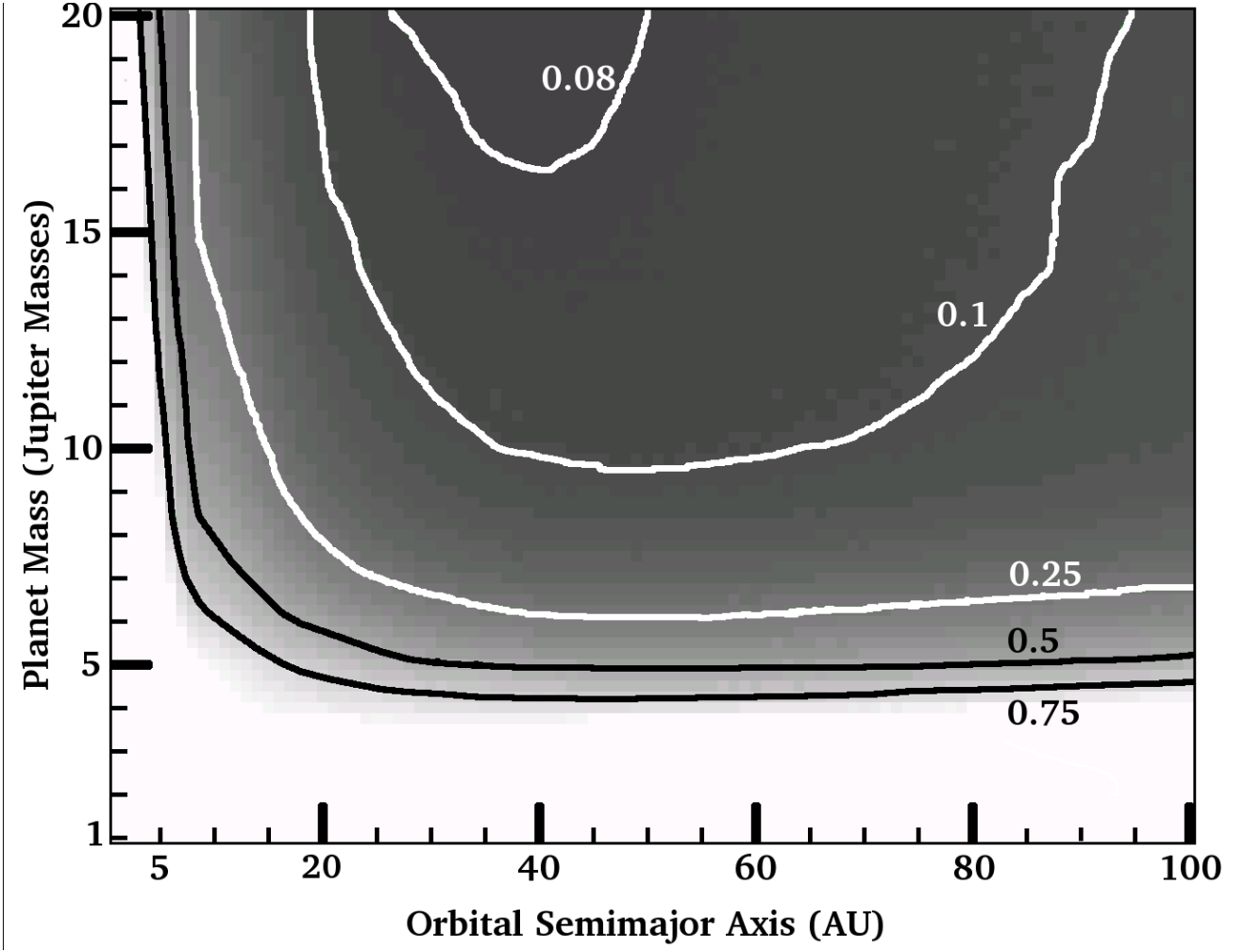}
\caption{90\% confidence level upper limits on the probability
$P(M,a)$ that a star like those in our survey will have
a planet of mass $M$ and semimajor axis $a$.  This plot shows,
for example, that our survey constrains the abundance
of 10 \mjup~or more massive planets with orbital semimajor 
axes between 22 and 100 AU to be less than 15\% around
sun-like stars.  The abundance of 5 \mjup~or more massive
planets between 25 and 94 AU is constrained to be less
than 50\%.  The latter range does not extend all the way to 
100 AU because our sensitivity
to planets in very distant orbits decreases somewhat
due to the possibility of their lying beyond our field
of view.
\label{fig:bmcgrid1}}
\end{figure}

Figure \ref{fig:bmcgrid1} shows the 90\% confidence upper limit on
$P(M,a)$ as a function of mass $M$ and semimajor axis $a$.
Each pixel represents thousands of realizations of our survey,
with $P(M,a)$ finely adjusted to reach the correct value.
Contours are overplotted showing where $P(M,a)$
is less than 8\%, 10\%, 25\%, 50\%, and 75\%, with 90\% confidence.
Note that $P(M,a)$, the value constrained by our
simulations, is a probability rather than a fixed fraction.  
The probability is the more scientifically interesting
number, but is harder to constrain.
For example, if 3.7\% is the fraction of \textit{the actual stars in our sample}
that have planets with easy-to-detect properties, there are
2 such planets represented in our 54-star survey.
However, if the \textit{probability} of a star
\textit{like those in our sample} having such a planet
is 3.7\%, there is still a nonzero probablity (13\% in this case) 
that no star in our sample actually has such a planet.

The results presented in Figure \ref{fig:bmcgrid1} can
be interpreted as model-independent constraints on planet
populations.  For example, with
90\% confidence we find that less than 50\% of
stars with properties like those in our survey have a 5 \mjup~or
more massive planet in an orbit with a semimajor axis between
30 and 94 AU.  Less than 25\% of stars like those
in our survey have a 7 \mjup~or more massive planet between
25 and 100 AU, less than 15\% have a 10 \mjup~or more massive 
planet between 22 and
100 AU, and less than 12\% have a 15 \mjup~or more massive planet/brown
dwarf between 15 and 100 AU.  Going to the most massive objects
considered in our simulations, we can set limits ranging inward
past 10 AU: we find that less than 25\% of 
stars like those surveyed have a 20 \mjup~object orbiting 
between 8 and 100 AU.  These constraints hold
independently of how planets are distributed in terms
of their masses and semimajor axes. 

HR 8799 appears to have a remarkable system of
three massive planets, seen at projected
distances of 24, 38, and 68 AU, with masses
of roughly 10, 10, and 7 \mjup, respectively \citep{hr8799}.
Using a Monte Carlo simulation like those used to
create Figure \ref{fig:bmcgrid1}, we find with
90\% confidence that less than 8.1\% of stars like
those in our survey have a clone of the HR 8799 planetary
system.  For purposes of this simulation we adopted the masses
above, and set the planets' orbital radii equal to their projected separations.
Our 8.1\% limit represents a step toward determining
whether or not systems of massive planets in wide orbits
are more common around more massive stars such as
HR 8799 than FGK stars such as those we have surveyed.

\subsection{Our Survey in the Big Picture} \label{sec:bigpic}

The surveys of \citet{kasper} and \citet{biller1},
have set constraints on the distributions
of extrasolar planets similar to those we present herein,
while \citet{nielsen} and especially \citet{GDPS} 
have set stronger constraints.  
More recent analyses by \citet{nielsenclose} and \citet{chauvin}
also provide constraints on the planetary distribution.  For
example, \citet{nielsenclose} provide a 68\% confidence that the 
\citet{cumming} distribution can be excluded for a truncation 
radius of 28 AU. However, if
different models are used this number jumps to 83 AU.  Chauvin et al.
indicate  a similar limit from analyzing their results using Baraffe
et al. 2003 models.  For the standard parameters they indicate a maximum
permitted truncation radius of approximately 35 AU.  In this context, 
the results presented here
provide looser constraints on the planet distribution, but provide an
independent check on the model-dependent systematic errors which may
exist with shorter wavelength data, due to incorrect model brightness
estimates or age determination.

Theoretical spectra of self-luminous extrasolar planets are
very poorly constrained observationally.  The recent detections
of possible planets around HR 8799 \citep{hr8799}, Fomalhaut \citep{fomalhaut},
and $\beta$ Pic \citep{betapic} are either single-band
($\beta$ Pic) or only beginning to be evaluated at multiple
wavelengths (HR 8799, Fomalhaut).  The candidate planets
orbiting HR 8799 and $\beta$ Pic are hotter than we would
expect to find orbiting middle-aged stars such as those
in our survey, while HST photometry of Fomalhaut b suggests
much of its brightness is starlight reflected from a
circumplanetary dust disk.  Our survey, and other
exoplanet surveys, must therefore
be interpreted using models of planetary spectra that are not yet
well-tested against observations.

Such models predict brightnesses in the $H$ band, and particularly
in narrow spectral windows within the $H$ band, that
are enormously in excess of black body fluxes.  The constraints
set by \citet{masciadri,biller1,GDPS,nielsen,nielsenclose}; and \citet{chauvin}
depend on the accuracy of these predictions of
remarkable brightness in the $H$ band.  The $L'$ and
$M$ bands that we have used are nearer the blackbody
peaks of low-temperature self-luminous planets, and might
be expected to be more reliable.  

However, \citet{L07} and \citet{reid} 
suggest that the $M$ band brightness at least of hotter extrasolar
planets will be less than predicted by \citet{bur} due to above-equilibrium
concentrations of CO from convective mixing.  \citet{NCE} present
new models indicating the effect is present for planets with
\Teff~ranging from 600 to 1800K.  The maximum $M$ band flux
supression is about 40\%, and flux supression disappears 
completely for \Teff~below 500K.  Based on \citet{bur},
this \Teff~value corresponds to planets of about 3.5,
6.5, 12, and 15 \mjup~at ages of 100 Myr, 300 Myr, 1 Gyr,
and 2 Gyr, respectively.  In many cases our $M$ band observations 
were sensitive to planets at lower masses than these values, 
and therefore \Teff~lower than 500K, implying that the CO flux
supression would have no effect on our mass limits.  In other
cases our $M$ band sensitivity did not extend so low.  However,
given that $M$ band observations formed a relatively small part
of our survey, and CO supression would affect only a fraction
even of them, the total effect on the statistical conclusions
of our survey should be entirely negligible.  

Theoretical spectra such as those of \citet{bur} may or
may not be more reliable in the $L'$ and $M$
bands than at shorter wavelengths.  However, so long as
the models remain poorly constrained by observations at
every wavelength, conclusions based on observations at
multiple wavelengths will be more secure.  Our survey,
with that of \citet{kasper}, has diversified planet imaging
surveys across a broader range of wavelengths.

In another sense our survey differs even from that of
\citet{kasper}: we have investigated older stars.
This is significant because planetary systems up to ages of 
several hundred Myr may still be undergoing substantial
dynamical evolution due to planet-planet interactions \citep{juric,levison}.  
Our survey did not necessarily probe the same planet
population as, for example, those of \citet{kasper} and \citet{chauvin}.

Finally, theoretical models of older planets are likely
more reliable than for younger ones, as these planets are further
from their unknown starting conditions and moving toward a well-understood,
stable configuration such as that of Jupiter.  It has been suggested by
\citet{faintJup} and \citet{fortney} that theoretical planet models 
such as those of \citet{bur} and \citet{bar} may overpredict 
the brightness of young ($<$ 100 Myr) planets 
by orders of magnitude, while for older planets the models are more accurate.

We have focused on nearby, mature star systems, and
have conservatively handled the ages of stars.
This makes our survey uniquely able to confirm that the rarity 
of giant planets at large separations around solar-type stars,
first noticed in surveys strongly weighted toward
young stars, persists at older system ages.  It is not an
artifact of model inaccuracy at young ages due to
unknown initial conditions.

\section{The Future of the $L'$ and $M$ Bands} \label{sec:long}
In the $L'$ and $M$ bands, the sky brightness is much worse than at
shorter wavelengths.  However, models (e.g., \citet{bur}) predict 
that in the $L'$ and $M$ bands, planets fade less severely with 
increasing age (or, equivalently, decreasing \Teff).  
Also, planet/star flux ratios are more favorable in the $L'$ and 
$M$ bands than at shorter wavelengths such as the $H$ and $K_S$ bands.  

It makes sense to use the $L'$ and $M$ bands on bright stars,
where the planet/star
flux ratio is a more limiting factor than the sky brightness.
In \citet{newvega}, we have shown that $M$ band observations
tend to do better than those at shorter wavelengths
at small separations from bright stars.  

The $L'$ and $M$ bands are most useful, however, for detecting 
the lowest temperature planets, which have the reddest 
$H - L'$ and $H - M$ colors.  Such very low temperature
planets can only be detected around the nearest stars, so
it is for very nearby stars that $L'$ and $M$ band observations
are most useful.  For distant stars, around which 
only relativly high \Teff~planets can be detected,
the $H$ and $K_S$ bands are much better.  We will now quantitatively
describe the advantage of $L'$ and $M$ band observations over shorter
wavelengths for planet-search observations of nearby stars.

Most AO planet searches to date have used the $H$ and $K_S$ bands, 
or specialized filters in the same wavelength regime.  
While the $K_S$ band has been used extensively to search for 
planets around young stars \citep{masciadri,chauvin}, our comparison 
here will focus on the $H$ band regime.  Models indicate it offers
better sensitivity than the $K_S$ band except for planets
younger than 100 Myr \citep{bur,bar}, and most of the stars
we will suggest the $L'$ and the $M$ bands are useful for
will be older than this. The most sensitive $H$-regime planet search
observations made to date are those of \citet{GDPS}, in part because
of their optimized narrow-band filter.  They attained an effective
background-limited point-source sensitivity of about $H=23.0$.
Based on the models of \citet{bur}, \citet{GDPS} would have set 
better planetary
mass limits than our observations around all of our own survey
targets except the very nearest objects, such as $\epsilon$
Eri and 61 Cyg.  Thus, at present, the $H$-regime delivers far better
planet detection prospects than the $L'$ and $M$ bands for most stars.

\begin{figure*}
\plottwo{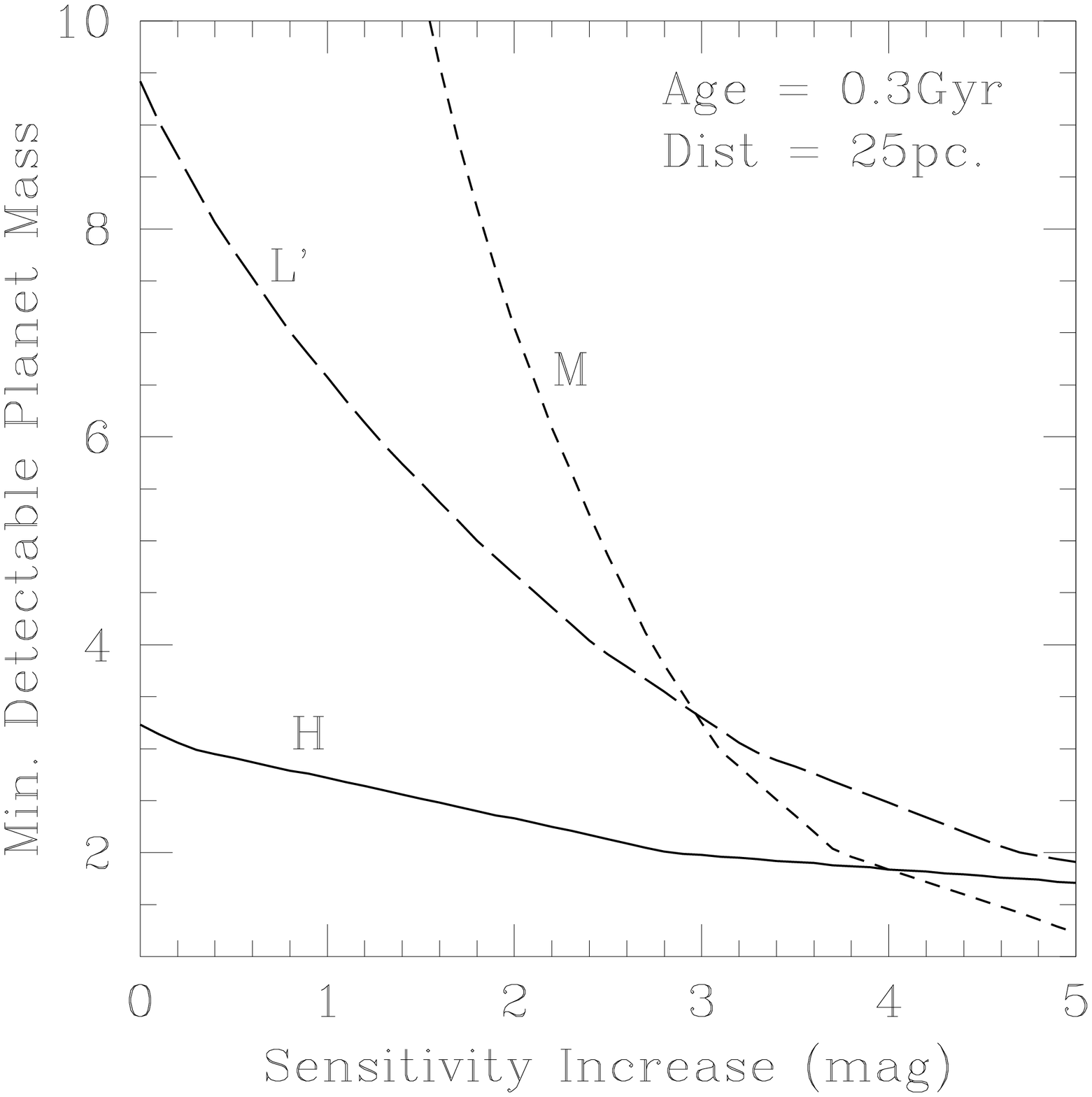}{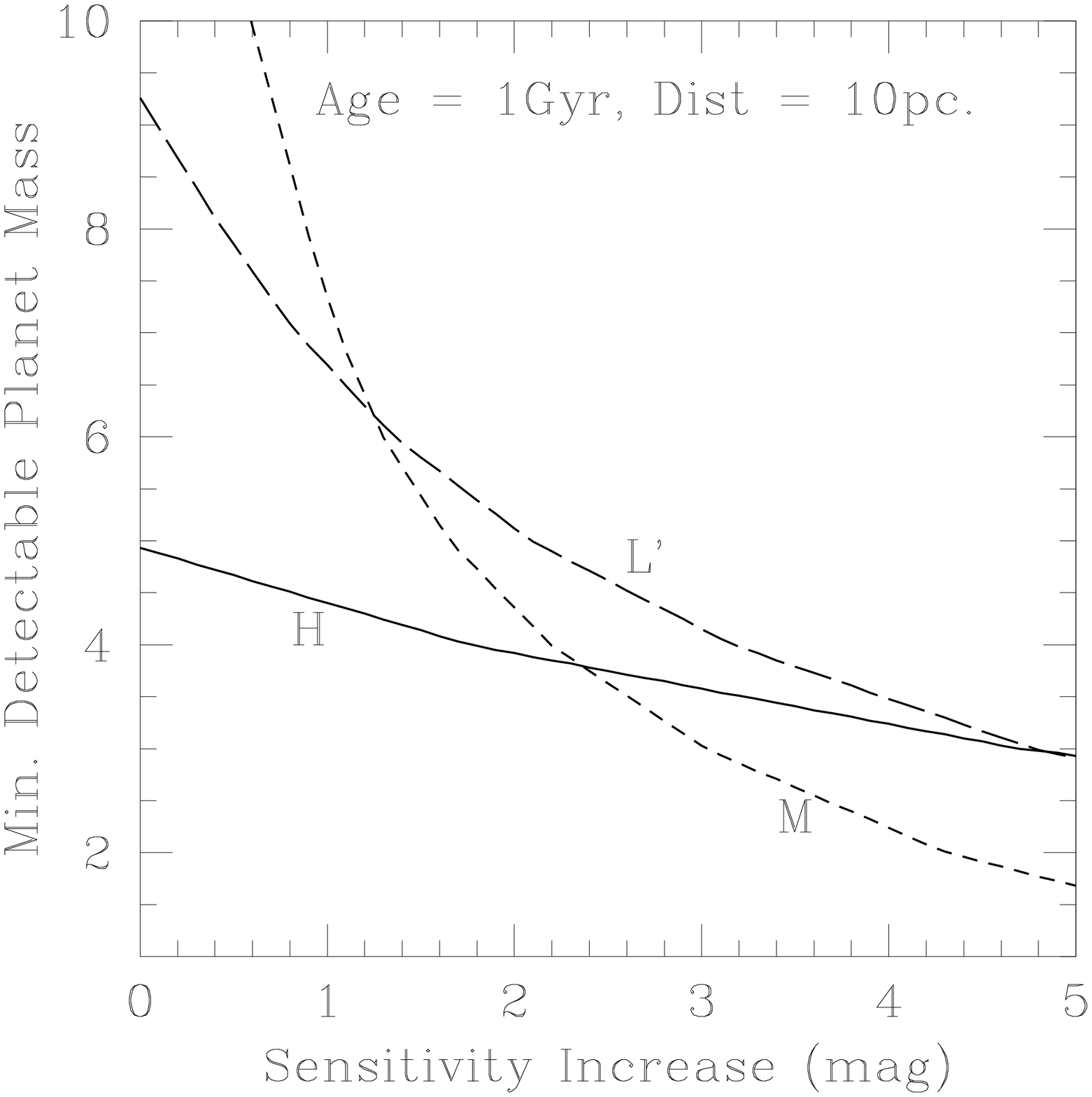}
\caption[$H$, $L'$, and $M$ band Compared]{Minimum detectable
planet mass in units of \mjup~for stars at 25pc (left) and 10pc (right),
in the $H$, $L'$, and $M$ bands, as a function of increase
over current sensitivity.  We have taken current sensitivities
to be $H = 23.0$, $L' = 16.5$, and $M = 13.5$.
While the $H$ band will likely remain the wavelength of choice
for planet search observations of stars at 25 pc and beyond,
an increase of only 2.4 mag over current sensitivities, even though
paralleled by an equal increase in $H$ band sensitivity, will render the 
$M$ band more sensitive than $H$ for planets around all stars
nearer than 10 pc.  The relative effectiveness of different
wavelengths depends sensitively on the distance to a star system,
but it is essentially independent of the stellar age, as explained
in the text.
\label{fig:HLM1}}
\end{figure*}

\begin{figure*}
\plotone{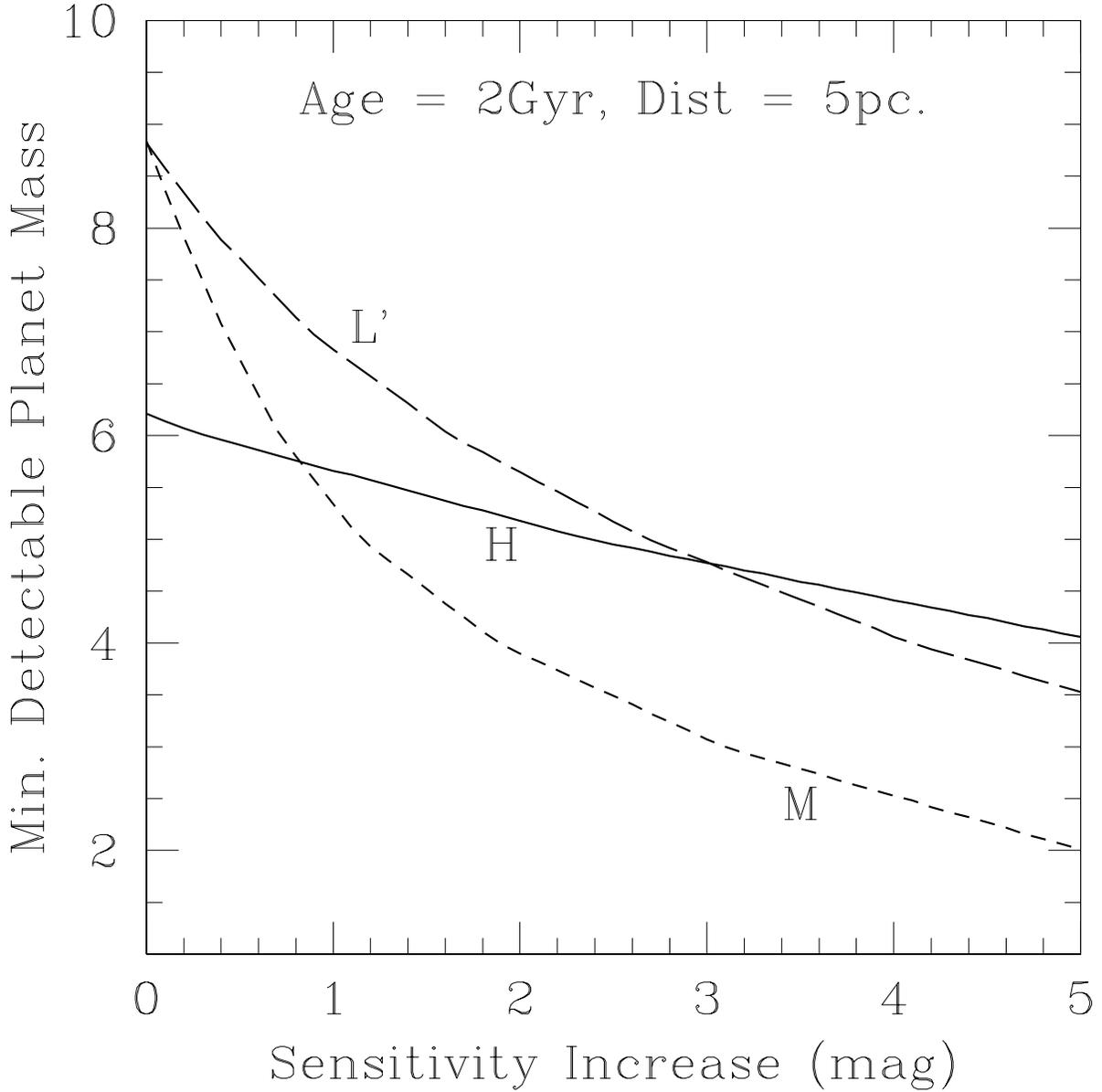}
\caption[$H$, $L'$, and $M$ band Compared]{Minimum detectable
planet mass in units of \mjup~for stars at 5pc, in the 
$H$, $L'$, and $M$ bands, as a function of increase
over current sensitivity.  We have taken current sensitivities
to be $H = 23.0$, $L' = 16.5$, and $M = 13.5$.  Given only
a 1 magnitude increase in $M$ band sensitivity, paralleled
by an equal increase at $H$ band, the $M$ band would be the
best wavelength for planet search observations around all
stars nearer than 5 pc.  While the sensitivity increases 
required to render $M$ preferable in Figure \ref{fig:HLM1}
require substantial improvements to existing instruments
and telescopes, the 1 mag increase required at 5 pc could
be obtained by simply increasing the exposure time.
As with Figure \ref{fig:HLM1}, this result concerning the
relative effectiveness of different wavelengths is independent
of stellar age, to first order.
\label{fig:HLM2}}
\end{figure*}

However, as detector technology improves, larger telescopes are 
built, and longer
planet detection exposures are attempted, the sensitivity
at all wavelengths will increase.  This means that low-temperature 
planets, with their red IR colors, will be detectable 
at larger distances, and the utility
of the $L'$ and especially the $M$ bands will increase.
In Figure \ref{fig:HLM1} we show the minimum detectable planet
mass for hypothetical stars at 10 and 25 pc distance as a function of the
increase over current sensitivity in the $H$, $L'$, and $M$
bands, and in Figure \ref{fig:HLM2} we present the same comparison
for a star at 5 pc.  We have taken current sensitivity to be $H = 23.0$
(i.e., \citet{GDPS}), $L' = 16.5$, and $M = 13.5$ (i.e., the
present work, scaled to an 8m telescope such as \citet{GDPS}
used).  These are background limits, not applicable close
to bright stars.  Based on \citet{newvega}, we believe
the $L'$ and $M$ bands will do even better relative to $H$
closer to the star where observations are no longer
background limited. Of course $H$ band observations with
next-generation extreme AO systems such as GPI and SPHERE will offer improved
performance close to the star, but advances in $M$-band AO
coronography (e.g. \citet{phaseplate}), will also improve the
longer-wavelength results.  In any case, Figures \ref{fig:HLM1}
and \ref{fig:HLM2} compare background-limited performance only.

The supression of flux in the $M$ band
due to elevated levels of CO \citep{L07,reid} does
not apply to planets at the low temperatures relevant for
Figures \ref{fig:HLM1} and \ref{fig:HLM2}.
Based on \citet{bur}, the entire mass range covered by both 
Figures corresponds to planets with \Teff~below 500K, except for
planets with masses above 6.5 \mjup~in the left panel of
Figure \ref{fig:HLM1} (25 pc distance, 300 Myr age).  This
upper section of the 25 pc, 300 Myr panel is irrelevant to 
the important implications of the figure.  According to \citet{NCE}, 
there is no supression of the $M$ band for effective temperatures
below 500K.

We have deliberately chosen the characteristics of the hypothetical 
stars in Figures \ref{fig:HLM1} and \ref{fig:HLM2} to be 
less good than the best
available planet search candidates, so that in each case
stars closer and/or younger than the example
actually exist. Using the very youngest stars would also
have resulted in sensitivities better than 1 \mjup, a mass
regime not covered by the \citet{bur} models used in the Figures.

Figures \ref{fig:HLM1} and \ref{fig:HLM2} illustrate 
three very important points.
First, the $L'$ band appears to have only secondary usefulness since
either the $H$ band or the $M$ band always offers sensitivity
to lower-mass planets.
Second, Figure \ref{fig:HLM2} shows that with a relatively 
minor increase of 1 magnitude in sensitivity, 
the $M$ band will be sensitive to lower-mass 
planets around all stars within 5 pc than can be
detected with $H$ band observations, even if the $H$ band sensitivity
increases the same amount. Third, Figure \ref{fig:HLM1}
shows that the advantage of the $M$ band decreases with
increasing distance, but that as larger telescopes and
longer exposures increase sensitivities to 2.5 mag
above present levels, the $M$ band will be superior
to $H$ out to 10 pc.  With an increase of 4 mag, the $M$ 
band would surpass $H$ out to 25 pc -- but as such a large
sensitivity increase would be difficult to achieve, $H$ band
will likely remain the primary wavelength for stars at
25 pc and beyond.  For stars closer than 10 pc, however,
the $M$ band already offers excellent sensitivity that
has barely been exploited so far.  Given reasonable
sensitivity increases, $M$ should become the primary
band for planet searches around stars at a distance
of 10 pc or less.

Interestingly, the conclusions of Figures
\ref{fig:HLM1} and \ref{fig:HLM2} are essentially independent 
of age: extensive calculations by \citet{thesis} showed that 
the relative usefulness of 
different wavelengths had only a weak dependence on age, for
stars at a fixed distance -- and even this weak age dependence
could change sign on switching from the models of \citet{bur} to 
those of \citet{bar}.  This means that
if we change the ages of the stars in Figures \ref{fig:HLM1} and \ref{fig:HLM2}
but leave the distances the same, the $L'$, $M$, and $H$ band
curves will slide up or down but remain essentially fixed
in their relative positions.  For example, given a 3 magnitude increase
in sensitivity at both wavelengths, $M$ band observations
will detect lower mass planets than $H$-band ones around
a star at 10 pc, whether the stellar age is 5 Gyr, 1 Gyr,
or 100 Myr.  This is to be expected, since if one dials down
the age of a given hypothetical star system, the \Teff~(and
therefore IR color) of the faintest detectable planets will 
remain about the same, though their masses will decrease.

Again, Figures \ref{fig:HLM1} and \ref{fig:HLM2} apply only to 
background-limited
sensitivity.  However, given the much more favorable planet/star flux
ratios in the $M$ band relative to $H$, we would expect the longer
wavelength observations to remain equally competitive closer to
the star.  Advances in $M$ band coronography will likely parallel
the development of $H$ band extreme AO systems such as GPI and
SPHERE.  Though at present they are
surpassed in sensitivity by $H$-regime observations for
all but the nearest stars, the $L'$ and especially the $M$ bands
hold considerable promise for the future.

\section{Conclusion} \label{sec:concl}
We have surveyed unusually nearby, mature star systems
for extrasolar planets in the $L'$ and $M$ bands using
the Clio camera with the MMT AO system.  By extensive
use of blind sensitivity tests involving fake planets
inserted into our raw data (reported in detail in
\citet{obspaper}), we 
established a definitive significance vs. completeness relation for planets in
our data, which we then used in Monte Carlo simulations
to constrain planet distributions.

We set interesting limits on the masses of planets and
brown dwarfs in the star systems we surveyed, but
we did not detect any planets.  Based on this
null result, we place constraints on the
power laws that may describe the distribution of
extrasolar planets in mass and semimajor axis.
We also place constraints on planet abundances
independent of the distributions. If the distribution of planets
is a power law with $dN \propto M^{\alpha} a^{\beta} dM da$, the
work of \citet{cumming} and \citet{butlercat} indicates that the 
most optimistic (i.e. planet-rich) case permitted by the statistics of known
RV planets correponds to about $\alpha = -1.1$ and $\beta = -0.46$.
Normalizing the distribution to be consistent with RV 
statistics, we find that these values of $\alpha$ and
$\beta$ are ruled out at the 90\% confidence level, unless the
semimajor axis distribution is truncated at a radius $R_{trunc}$
less than 110 AU.  Though $\beta=0.0$ is not physically plausible, 
previous work has sometimes used it an example: 
for $\alpha=-1.31$, corresponding
to the best-fit value from \citet{cumming},
we rule out $\beta=0.0$ unless $R_{trunc}$ is less than 38 AU.
Independent of distribution models,
with 90\% confidence no more than 50\% of stars like those in
our survey have a 5 \mjup~or more massive planet orbiting
between 30 and 94 AU, no more than 15\% have a 10 \mjup~planet
orbiting between 22 and 100 AU, and no more than 25\% have a
20 \mjup~object orbiting between 8 and 100 AU.

Our constraints on planet abundances are similar to those placed by
\citet{kasper} and \citet{biller1}, but less tight
than those of \citet{nielsen} and especially \citet{GDPS},
The recent work of \citet{nielsenclose} and \citet{chauvin}
also placed tighter constraints on exoplanet distributions than
our survey.  However, we have surveyed a more nearby, older set of
stars than any previous survey, and have therefore
placed constraints on a more mature population of planets.
Also, we have confirmed that a paucity of giant planets
at large separations from sun-like stars is robustly observed at a wide
range of wavelengths.

The best current $H$ regime observations, those of \citet{GDPS},
would attain sensitivity to lower mass planets than did our
$L'$ and $M$ band observations for all of our survey targets
except those lying within 4 pc of the Sun.  However, as larger
telescopes are built and longer exposures are attempted, the
sensitivity of $M$ band observations may be expected to increase
at least as fast as that of $H$ band observations (in part because
$M$ band detectors are currently a less mature technology).
As shown in Figures \ref{fig:HLM1} and \ref{fig:HLM2}, a modest increase from 
current sensitivity levels, even
if paralleled by an equal increase in $H$ band sensitivity, would render
the $M$ band the wavelength of choice for extrasolar planet
searches around a large number of nearby stars.

\section{Acknowledgements} 
This research has made use of the SIMBAD online database,
operated at CDS, Strasbourg, France, and
the VizieR online database (see \citet{vizier}).

We have also made extensive use of information and code
from \citet{nrc}. 

We have used digitized images from the Palomar Sky Survey 
(available from \url{http://stdatu.stsci.edu/cgi-bin/dss\_form}),
 which were produced at the Space 
Telescope Science Institute under U.S. Government grant NAG W-2166. 
The images of these surveys are based on photographic data obtained 
using the Oschin Schmidt Telescope on Palomar Mountain and the UK Schmidt Telescope.

Facilities: \facility{MMT, SO: Kuiper}

\end{document}